# A novel approach to the synthesis of nanostructured metal-organic films: X-ray radiolysis of silver ions using a Langmuir monolayer as a template.

*François Muller[1,2], Philippe Fontaine[1\*], Samy Remita[3], Marie-Claude Fauré[2\*], Emmanuelle Lacaze[4] and Michel Goldmann[1,2\*]*

Laboratoire pour l'Utilisation du Rayonnement Electromagnétique, CNRS UMR 130, BP 34, 91898 Orsay Cedex, France; Objets Complexes et Interfaces d'Intérêt Biologique, CNRS FRE 2303, université Paris 5, 45 rue des Saint-Pères, 75270 Paris Cedex 06, France; Laboratoire de Chimie et Biochimie Pharmacologiques et Toxicologiques, CNRS UMR 8601, université Paris 5, 45 rue des Saint-Pères, 75270 Paris Cedex 06, France ; Groupe de Physique des Solides, CNRS UMR 7588, université Paris 6, 2 place Jussieu 75251 Paris Cedex 05 France ;

Corresponding author: philippe.fontaine@lure.u-psud.fr

RUNAWAY TITLE: X-ray radiolysis of silver ions using a Langmuir monolayer as a template.

[1]Laboratoire pour l'Utilisation du Rayonnement Electromagnétique (LURE).
[2]Objets Complexes et Interfaces d'Intérêt Biologique (OCIIB).
[3]Laboratoire de Chimie et Biochimie Pharmacologiques et Toxicologiques (LCBPT).
[4]Groupe de Physique des Solides (GPS)

[\*] Now at Groupe de Physique du Solide

**Abstract**

An application of the radiolysis method using X-ray synchrotron beam is developed as a novel approach to the synthesis of metal-organic films with controlled shapes and thickness. We demonstrate that a Langmuir monolayer deposited onto a silver ion-containing subphase, irradiated by an incident beam impinging below the critical angle for total reflection, induces the synthesis of a stable nanostructured silver-organic ultrathin film at the air-water interface. The X-ray scattering is also used to monitor *in situ* the structure of the silver layer during the synthesis process. The layer is observed by AFM after its transfer onto a silicon substrate. One observes a film thickness of 4.6 nm, in good agreement with the X-ray penetration depth, about 4.5 nm. The silver structure is oriented by the initial organic film phase. This experiment demonstrates the considerable potential of this approach to produce various controlled metal-organic films with a surfactant self-assembly as a template.

**Introduction**

Synthesis of metal nanoclusters and ultrathin metal-organic interfaces with tunable properties (optical, magnetic, electronic, catalytic, …) into specific physicochemical environments becomes a fascinating challenge from a fundamental as well as an applied point of view[1,2]. These materials are usually generated by the chemical reduction of metal ion precursors. Among the possibilities, radiolysis method is efficient to control this synthesis in solution and allows formation of mono- and multimetal clusters with an homogeneous distribution[3]. Irradiation of aqueous solutions containing metal ion precursors induces metal ion reduction into atoms which coalesce and finally form metal aggregates. This reaction can operate under various conditions. For instance, metal clusters[2-4] and monodisperse metal colloids in presence of ligands (ethylenediaminetetraacetic acid EDTA, Triton X-100[5]) or of polymers (polyvinyl alcohol PVA, polyacrylate PA[6]) have been successfully radio-synthesized. However, from our knowledge, such a procedure has not been attempted for the synthesis of metal-organic interfaces either with self-assembled surfactant systems in aqueous solutions (spheres, cylinders…) or with Langmuir monolayers (insoluble monomolecular surfactant layers) as templates. Indeed, due to the interaction between the polar headgroups of anionic (cationic) surfactant molecules and cationic (anionic) metal ions,[7] the radiolysis method should provide an atom coalescence localized onto the surface of a direct surfactant self-assembly. Owing to the diversity in surfactant phase diagrams, this approach presents a considerable potential toward the synthesis of metal-organic interfaces with numerous controlled shapes.

In this letter, we demonstrate that the irradiation of the surface of a Langmuir monolayer in condensed state deposited onto a metal ion-containing subphase leads to the formation of a metal-organic layer oriented by the organic monolayer. We develop a simple and versatile variation of the radiolysis method using a well-collimated X-ray beam in grazing incidence geometry as radiation source. In this geometry, the incident X-ray beam strikes the interface at an angle below the critical angle for total external reflection and the X-ray penetration in the subphase is restrained to a thin layer, the thickness of which is defined by the grazing incidence angle $\alpha_i$ (Figure 1). The reduction of metal ions is then initiated in the vicinity of the polar headgroups and templated by the initial organic structure.

Silver ions as metal precursors and behenic acid as organic surfactant have been chosen for this first study since silver reduction properties and behenic acid Langmuir films have been extensively studied.[8,9] Moreover, our choice of silver is also directed by the huge activity field in silver colloids and surfaces due to their ability in catalysis, their application in photochemistry process and also by the use of silver as substrates for surface enhanced Raman spectroscopy (SERS)[10].

**Experimental section**

**Sample preparation**. Silver sulfate $Ag_2SO_4$ (Sigma, purity >99%) is dissolved in ultrapure water (Millipore system, 18 MΩ.cm) at a silver ion concentration of 1.5 x $10^{-4}$ Mol/L and sheltered from light due to its photosensitivity. 80 µl of a 5 $10^{-3}$ Mol.L$^{-1}$ solution of behenic acid ($C_{21}H_{43}COOH$, Sigma) in pure chloroform ($CHCl_3$, Prolabo Normapur grade) is spread onto the silver ion-containing solution in a PTFE Langmuir through of 700 cm$^2$ area. It leads to a molar ratio between silver cations and organic molecules of $Ag^+/C_{21}H_{43}COOH$ = 375. The temperature of the subphase is maintained at 22 ± 0.5°C and the monolayer slightly compress up to π=15 mN/m surface pressure using a PTFE barrier. The Langmuir through is enclosed in a gas tight box flush by a continuous helium flux (Figure 1). It allows the deoxygenation of the aqueous subphase in the experiment time scale and prevents the bulk oxygenation by air during the radiolytical process.

**Details on radiolysis method of silver ion-containing solutions**. Irradiation of a deoxygenated aqueous solution containing silver ion precursors produces, by radiation-induced reduction, atoms which coalesce and finally form silver aggregates. Radiolysis of such a solution ensures at first the homogeneous distribution of free radicals formed by ionization and excitation of water molecules which are the most important species[11]:

$$H_2O \rightarrow e_{aq}^{-}, H^{+}, H^{\bullet}, OH^{\bullet}, H_2O_2, H_2 \quad (1)$$

Generated $e_{aq}^{-}$ free radicals are strong reductive agents and reduce easily silver cations into atoms:

$$Ag^{+} + e_{aq}^{-} \rightarrow Ag^{o} \quad (2)$$

Generated $OH^{\bullet}$ free radicals are strong oxidative agents which may counterbalance the metal reduction. Primary or secondary alcohols are usually added in excess to the solutions to scavenge $OH^{\bullet}$ free radicals (and also $H^{\bullet}$ free radicals) during radiolysis. Absolute ethanol ($CH_3CH_2OH$, Prolabo Normapur grade) was then added to the silver ion-containing subphase (see above) at a concentration of $10^{-2}$ mol.L$^{-1}$ to scavenge $OH^{\bullet}$ free radicals[12]. It generates $CH_3^{\bullet}CHOH$ free radicals which are able to reduce silver ions. Silver atoms and then aggregates are formed with a homogeneous distribution and a well known reduction yield throughout the solution following a multi-step process:

$$Ag_n^{x+} + Ag_m^{y+} \rightarrow Ag_{n+m}^{(x+y)+} \quad (3)$$

**Irradiation geometry**. The surface was irradiated with X-ray radiation under grazing incidence using the D41B beamline at the LURE synchrotron source (Orsay, France). The X-ray wavelength, λ=0.1605 nm (7.7 keV), was selected using a focusing Ge(111) crystal. The incidence angle was fixed for all the experiments at $\alpha_i$=2 mrad using a mirror, leading to an X-ray penetration depth inside the aqueous subphase of about 4.5 nm[15]. The footprint dimension, 75 x 1 mm$^2$ is defined by the deflecting mirror length and the incident beam width.

**Interface characterization**. In the present study, the surface is characterized through X-ray scattering to determine *in situ* and in real time the structural changes of the interface during radiolysis[13]. The thickness and the structure of the interface as a function of the irradiation time was investigated using, respectively, grazing incidence X-ray surface scattering[15] and grazing incidence X-ray diffraction[15]. The helium flux (Figure 1), mandatory for deoxygenation of the solution, also reduces the scattering by air. The intensity of the X-ray beam diffracted by the interface was monitored using a vertical gas-filled (Ar-$CO_2$) position sensitive detector (PSD) as a function of the in-plane component $q_{xy}$ selected by means of a soller slit collimator. The resulting $q_{xy}$ resolution was 0.07 nm$^{-1}$. The intensity of the X-ray beam scattered by the interface was monitored using the PSD as a function of $\theta_z$ the vertical angle from the interface. When the radiolysis process is completed (when the X-ray scattering spectra does not present any more evolution), the surface layer is transferred onto a silicon substrate. For this purpose, substrates (silicon wafers covered by the native $SiO_2$ layer) were previously immersed in the water subphase below the X-ray footprint, at about 1 mm depth. Transfer is obtained by removing slowly the water from the through down to the situation where the water level is below the substrate surface, depositing then the layer onto the silicon surface. The sample is then dried by natural evaporation. Images of the surface are then obtained using an Atomic Force Microscope (AFM, nanoscope 3100, Veeco) in tapping mode.

**Results and Discussion**

The normal structure of the interface was investigated using grazing incidence X-ray surface scattering[14]. In the absence of silver cations in the subphase, the surface scattering spectrum is almost featureless due to the low contrast between water and behenic acid monolayer (inset of Figure 2). The shape of the curve is only determined by the height fluctuations of the interfaces[16]. As radiolysis proceeds (with silver cations), one observes the appearance of oscillations in the surface scattering spectrum (Figure 2). Such oscillations are the signature of the appearance of a new layer at the interface[14,16]. They indicate the progressive synthesis of a silver layer under the interface. Adjustment of these spectra with theoretical curves will be presented in detail in a further publication.

After about 12 hours of irradiation, no more change in the spectrum is observed (Figure 2). It indicates the end of the layer growth on the depth of the X-ray beam, probing that the silver ions reduction reaches saturation in the overall irradiated volume. From this moment onward, no more layer growth due to radiolysis happens. If a layer growth still proceeds, it might be only ensured by the silver layer itself in an electrochemical sense. However, X-ray photons with an energy around 8 keV are not able to create an adequate surface potential on the silver layer to induce such an electrochemical growth. This experiment also probes that the synthesized layer is stable inside the overall irradiated volume and tethered to the interface by the Langmuir film. We note that the typical yellow-brown coloration of the solutions, signature of the presence of silver colloids in the bulk, has not been observed indicating that the atom coalescence occurs only in the vicinity of the surface.

The in-plane structure of the interface was investigated by using grazing incidence X-ray diffraction[16]. In the absence of silver cations in the subphase, one obtains the typical diffraction spectra (2 peaks at about 1.46 Å$^{-1}$ and 1.47 Å$^{-1}$, Figure 3a) expected for a behenic acid monolayer at π=15 mN/m (L$_2$ phase, next-neighbour tilt)[9]. However, as the radiolysis proceeds (with silver cations), the diffraction patterns after about 1 hour of irradiation reveal the growth of new peaks at $q_{xy}$=1.37 Å$^{-1}$ and at $q_{xy}$=1.58 Å$^{-1}$ (Figure 3a). Intensity of the initial L$_2$ phase signal progressively decreases while this new structure grows up. A larger exploration of the reciprocal space shows the appearance of higher order diffraction peaks. Irradiation of a silver ion-containing subphase without behenic acid monolayer onto the surface does not lead to any structure (results not shown) proving that the organic monolayer is necessary to tether the silver atoms at the interface.

The intensity and the sharpness both in-plane and out-of-plane of the new peaks shown in Figures 3a, 3b and 3c confirm that the new structure does not correspond to the organic layer but to a silver layer localized under the surface. From their extension along $q_z$, one obtains a thickness of about 50 Å. This result also shows that the silver layer is organized and oriented. The ($q_{xy}$, $q_z$) intensity distributions (contour plots in Figure 3b and 3c) indicate that the silver crystals growth is oriented by the air-water interface. Indeed, a homogeneous distribution like powder or colloid distribution would lead to diffraction rings and not to separate peaks as observed on the contour plots (Figure 3b and 3c).

An AFM image of the film deposited onto silicon substrate is presented on figure 4. One clearly observes sharp domains of a few microns size. The domain thickness, 4.6 nm, is in very good agreement with the estimated 4.5 nm penetration of the X-ray evanescent wave. The angular shape of the domain is also in agreement with a crystallized structure. From this image, one can conclude that we indeed succeed in forming a silver layer by the radiolysis process. One observes regions which present a double thickness, about 9 nm, corresponding to the superposition of different domains. Indeed, from this image, it appears clearly that the layer has been broken, probably during the deposition, indicating that one should develop a "softer" transfert method. Note that the organic monolayer expected on top of the silver crystal is not clearly observed in this image, may be due to penetration of the AFM tip.

**Conclusion**

In conclusion, we demonstrate that the radiolysis approach using a grazing incidence X-ray beam allows to generate a stable nanostructured metal-organic film oriented by the air-water interface. The new radiolysis geometry provides a convenient and unusual tool for the *in situ* observation of both metal film structure and growth kinetic. It supplies a considerable extension to understand metal film synthesis as a function of metal ion concentration, pH-subphase, metal nature and surfactant molecules. It should lead to a better control of the metal-organic thin film synthesis using the radiolysis method. This versatile experiment also demonstrates the considerable potential of the radiolysis approach in the presence of self-assembled surfactant systems as templates. It is the first step toward the production of more and more complex metal-organic systems with controlled shapes, structures and properties. A natural extension of the present work is the development of a method for a well characterization of the synthesized ultrathin layer with near field microscopies as well as atomic force microscopy (AFM) and scanning tunneling microscopy (STM). Moreover, using the same radiolysis approach onto a Langmuir monolayer in small separate domains as template is attractive to the synthesis of metal-organic nano-objects. Both works are currently in progress and will be reported in due course.

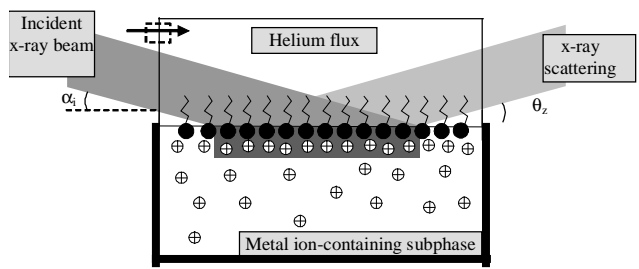

**Figure 1.** Schematic experimental geometry. Counterions and free radical scavenger (see text) were omitted for clarity.

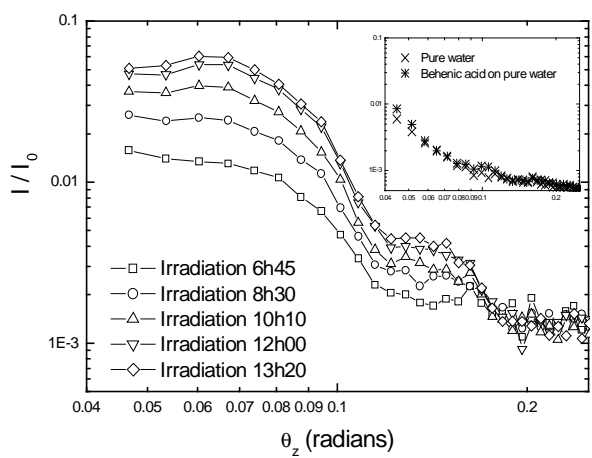

**Figure 2.** Grazing incidence x-ray surface scattering spectrum (normalized off-specular intensity $I/I_0$ as a function of the out-of-plane angle $\theta_z$, $\alpha_i=2$mrad, log-log representation) for a behenic acid monolayer at pH=7 and at $\pi=15$ mN/m on a silver ion-containing subphase (0.15 mmol.L$^{-1}$). In inset, same representation for pure water and for a behenic acid monolayer at pH=7 and at $\pi=15$ mN/m on pure water subphase.

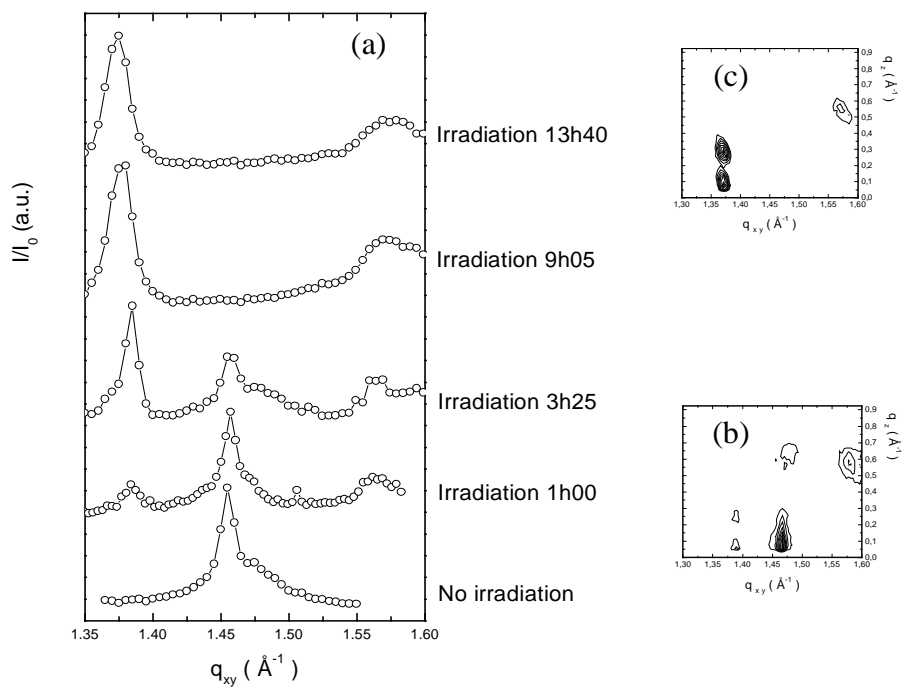

**Figure 3.** Grazing incidence x-ray diffraction as a function of the in-plane ($q_{xy}$) component of the scattering wave vector q during irradiation (a) and contour plots (Intensity as a function of in-plane, $q_{xy}$ and of the out-of-plane $q_z$ component) after 1h of irradiation (b) and after 13h40 of irradiation (c). The initial spectrum (no irradiation) has been obtained for a behenic acid monolayer at pH=7 and $\pi=15$ mN/m on pure water subphase and the others for a behenic acid monolayer at pH=7 and $\pi=15$ mN/m on a silver ions-containing subphase (1.5 10$^{-4}$ Mol/L). The two peaks at about 1.45 Å$^{-1}$ and 1.46 Å$^{-1}$ correspond to the structure observed for behenic acid chains in L$_2$ phase (next-neighbour tilt), peak at 1.375 Å$^{-1}$ and 1.58 Å$^{-1}$ correspond to the structure which grows during radiolysis.

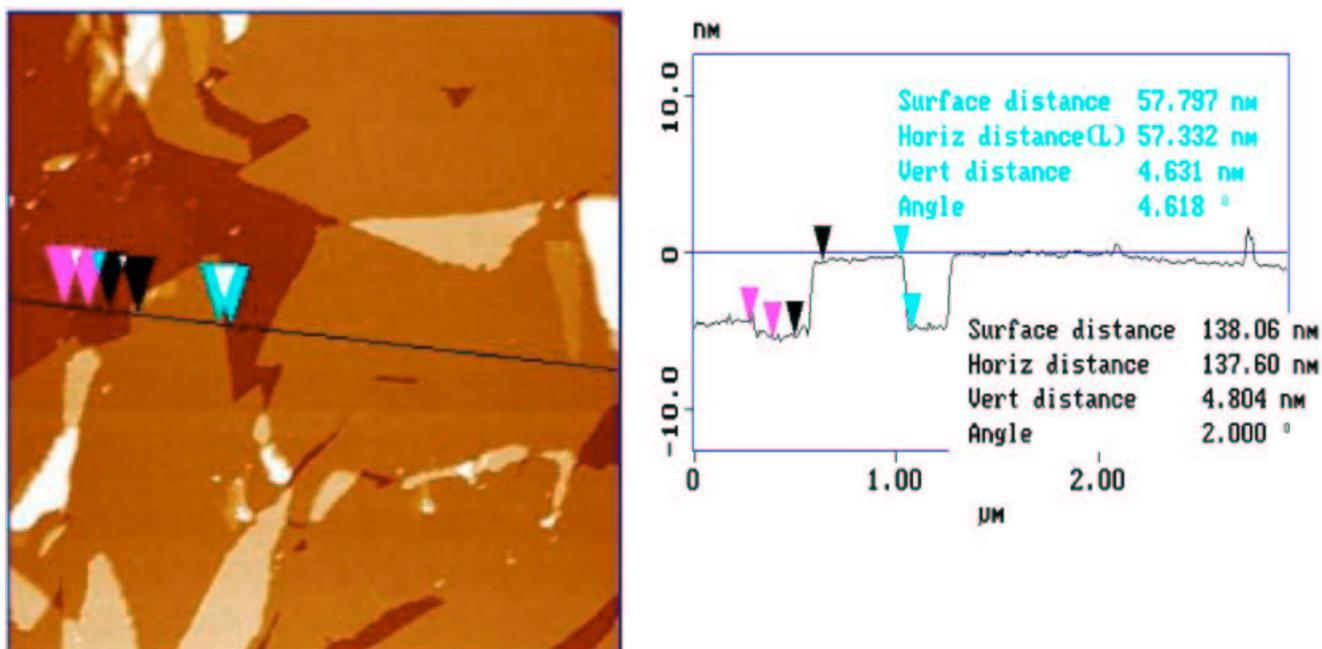

**Figure 4.** Atomic force microscopy image of the surface layer deposited on silicon wafer after complete radiolysis process at the air-water interface. One clearly observes sharp domains of a few microns size and of 4.6 nm thickness. These domains correspond to the silver layer formed by the surface irradiation which has been probably broken during the deposition process.